\documentclass{PoS}
\usepackage{epsfig}

\PoS{PoS(LAT2005)215}

\title{Low-energy J/$\psi$-Hadron Interactions from Quenched Lattice QCD }

\ShortTitle{Low-energy J/$\psi$-Hadron Interactions from Quenched Lattice QCD }

\author{\speaker{Kazuo Yokokawa}\\
        Department of Physics, University of Tokyo, Tokyo 113-0033, Japan\\
        E-mail: \email{yokokawa@nt.phys.s.u-tokyo.ac.jp}
        }

\author{Shoichi Sasaki\\
        RIKEN BNL Research Center, Brookhaven National Laboratory, NY 11973, USA\\
        E-mail: \email{ssasaki@phys.s.u-tokyo.ac.jp}
        }
\author{Arata Hayashigaki\\
        Institut f\" ur Theoretische Physik, J.W.~Goethe Universit\" at, D-60438 Frankfurt, Germany\\
        E-mail: \email{aratah@th.physik.uni-frankfurt.de}
        }
\author{Tetsuo Hatsuda\\
        Department of Physics, University of Tokyo, Tokyo 113-0033, Japan\\
        E-mail: \email{hatsuda@phys.s.u-tokyo.ac.jp}\\
        }

\abstract{
The J/$\psi$-hadron interaction is a key ingredient in
 analyzing the J/$\psi$ suppression in hot hadronic matter as well as  
 the propagation of J/$\psi$  in nuclei.
 As a first step to clarify the  J/$\psi$-hadron interactions at low energies,
 we have calculated  J/$\psi$-$\pi$, J/$\psi$-$\rho$ and J/$\psi$-nucleon 
scattering lengths by the quenched lattice QCD simulations with Wilson fermions 
for $\beta=6.2$ on $24^3\times48$ and $32^3\times48$ lattices.
Using the L\"uscher's method to extract the scattering length from the 
 simulations in a finite box, we find 
 an attractive interaction
 in the S-wave channel for all three systems:
  Among others, the J/$\psi$-nucleon interaction is most attractive.
 Possibility of the J/$\psi$-nucleon bound state is also discussed.
  }

\FullConference{XXIIIrd International Symposium on Lattice Field Theory\\
		 25-30 July 2005\\
		 Trinity College, Dublin, Ireland}

\begin{document}

\section{Introduction}
 A possible signature of the formation of the quark-gluon
 plasma (QGP) in relativistic heavy ion collisions
  is the suppression of J/$\psi$ due to color Debye screening  \cite{MS}.
 In the $Pb$-$Pb$ collisions at CERN-SPS, 
  the J/$\psi$ suppression beyond the normal nuclear absorption
 has been discovered \cite{NA50}. However, the data 
 may be described either by the color Debye screening due to deconfined
  quarks and gluons or
  by absorption/dissociation  due to comoving light hadrons.
 Recent BNL-RHIC data show a tendency that the  J/$\psi$ suppression is 
 almost independent of collision energies between 62 GeV and 200 GeV.
 The magnitude of the suppression is less than those predicted
 from color Debye screening or from the absorption by comovers, which 
  may be understood by the recombination of $c$ and $\bar{c}$ \cite{QM}.

To understand the mechanism of J/$\psi$ suppression in those experiments,
 we need more precise knowledge on the  J/$\psi$-hadron interactions.
In this report, 
 we show our recent results of J/$\psi$-hadron scattering lengths calculated
 in quenched lattice QCD simulations. (See \cite{AH,WE,TO,SV,SL} and references therein
  for other approaches.)
  
\section{Formulation}

Suppose we have two hadrons in a finite box. The effect of 
their interaction appears as an energy shift $\Delta E$ relative to the 
 non-interacting case. $\Delta E$ may be
  extracted from the correlator ratio $R(t)$ for large $t$;
\begin{eqnarray}
R(t) = \frac{G_{{\rm J}/\psi{\textit-}H}(t)}{G_{{\rm J}/\psi}(t)G_H(t)}  
\sim {\rm e}^{-(E-m_{{\rm J}/\psi}-m_H)\cdot t} = {\rm e}^{-\Delta E \cdot t},
\end{eqnarray}
where $G_{{\rm J}/\psi \textit{-}H}(t)$ and $G_{H}(t)$ are 
the J/$\psi$-hadron four-point function and the hadron two-point function, 
respectively.

In our calculation, we consider three scattering processes, 
J/$\psi$-$\pi$, J/$\psi$-$\rho$ and J/$\psi$-$N$(nucleon), 
which are most important for J/$\psi$ absorption by comoving hadrons 
in relativistic heavy ion collisions. 
Since each hadron has spins, 
we need to make spin projection to good total spin states  of J/$\psi$-hadron 
two body system.
For example,  the J/$\psi$-$N$ case reads
\begin{eqnarray}
G_{{\rm J}/\psi\textit{-}N}(t) = G_{1/2}(t)\hat{P}^{1/2}+G_{3/2}(t)\hat{P}^{3/2},
\end{eqnarray}
where $G_{1/2(3/2)}$ denotes
 the four-point function with a good spin quantum number ($J=1/2$ or 3/2)
 and $\hat{P}^{1/2(3/2)}$ is   the spin projection operator \cite{KSSS}.

The L\"uscher's formula tells us a relation between $\Delta E$ and 
the scattering observables such as the scattering length and the scattering phase shift.
 For the S-wave scattering phase shift, it reads
\cite{ML}
\begin{eqnarray}
 \tan\delta_0(q) = \frac{\pi^{3/2}\sqrt{q}}{Z_{00}(1,q)},
\label{tan-d}
\end{eqnarray}
with sign convention as which negative phase shift corresponds to repulsion.
Here $Z_{00}$ is a generalized zeta function defined by
\begin{eqnarray}
 Z_{00}(s,q)=\frac{1}{\sqrt{4\pi}}\sum_{\bf{n}}\frac{1}{(n^2-q)^s},\qquad
 q=\left( \frac{pL}{2\pi} \right)^2,
\end{eqnarray}
$p$ and $L$ are the relative momentum of the two hadrons and the spatial size of the box,
 respectively. Here it is assumed that the interaction range is finite.
 Outside the interaction range,
 the total energy of the system is related to the relative momentum $p$ as 
\begin{eqnarray}
\sqrt{m^2_{J/\psi}+p^2}+\sqrt{m_H^2+p^2} = E.
\end{eqnarray}
Here positive (negative) $p^2$ corresponds to the  repulsion (attraction).
 If there are no interactions between the hadrons, 
 $p$ takes discrete momentum in the finite box as 
 $p^2=(2\pi /L)^2\cdot n\;$ ($n=0,1,2,\cdots$).
 If there are interactions, $p$ receives an extra contribution $p_{\rm int}$
 and then a momentum squared divided by $(2\pi /L)^2$ is no longer an integer.

The S-wave scattering length is defined as 
$a_0\equiv \lim_{p\to 0} \tan(p)/p$ and is related to the zeta function
 through Eq.(\ref{tan-d}) as 
\begin{eqnarray}
 a_0 = \left. \frac{L\sqrt{\pi}}{2Z_{00}(1,q)} \right|_{n=0},
\label{a0-zeta}
\end{eqnarray}
where the scattering length is assigned to be negative (positive) for 
 repulsion (weak attraction). The hadron scattering 
 lengths based on the formulas Eqs.(\ref{tan-d}) and (\ref{a0-zeta})
  has been extensively studied for $\pi$-$\pi$ and $N$-$N$ systems
   in Refs.~\cite{Krms,SRB}.

 It is important here to discuss the asymptotic behavior of Eq.(\ref{a0-zeta})
 for large $L$ for the purpose of analyzing the system with attractive 
  interactions. The large $L$ expansion of the right hand side of Eq.(\ref{a0-zeta})
 at $q \sim 0$ leads to \cite{ML}
\begin{eqnarray}
\Delta E = -\frac{2\pi a_0}{M_{\rm res}L^3} \left( 
1+c_1 \left( \frac{a_0}{L} \right) +c_2 \left( \frac{a_0}{L} \right)^2
\right) +O(L^{-6}),
\label{largeL}
\end{eqnarray}
with $c_1=-2.837297$ and $c_2=6.375183$.
Let us try to solve Eq.(\ref{largeL}) in terms of $a_0$ for
 given $\Delta E$.
 In the case that $\Delta E >0$, both the expansion
  up to $O(L^{-4})$ and that up to $O(L^{-5})$ always have real and negative
   solutions. On the other hand, in the case that $\Delta E <0$,
  the expansion up to $O(L^{-4})$ gives no real solution
 for 
 \begin{eqnarray}
 \Delta E <  -\frac{\pi}{2|c_1|M_{\rm res}L^2},
\label{condition}
\end{eqnarray} 
although the expansion
  up to $O(L^{-5})$ always has a real solution.

  This observation implies that  some care must be taken to use the 
   expansion especially  for relatively strong attraction.
   Indeed, our lattice data show that the J/$\psi$ interaction with $\pi$, $\rho$
    and $N$ are all attractive and the condition
     Eq.(\ref{condition}) is met for J/$\psi$-$\rho$ and J/$\psi$-$N$ cases.
   Therefore, in our study,   
   we use Eq.(\ref{a0-zeta}) directly without the large $L$ expansion
   to extract $a_0$.
  
\section{Results}

In our simulation, we employed unimproved Wilson gauge action and Wilson 
 fermion. We have $\beta=6.2$ 
on $L^3\times T=24^3\times 48$ and $32^3\times48$ lattices with  
$\kappa$(charm) $=$ 0.1360 and $\kappa$(light) $=$ 0.1520, 0.1506, 0.1489.
In the physical unit, the lattice sizes are $L\sim$1.6, 2.1 fm,
 the lattice spacing is $a\sim$0.067 fm, and 
$m_{J/\psi}\sim$3.0 GeV and $m_\pi\sim$0.6-1.2 GeV.
The number of quenched gauge configurations  for smaller lattice
 is 161 and that for larger lattice is 169.
Our error estimates are all based on the Jackknife method.

%
%
\begin{figure}
\epsfig{file=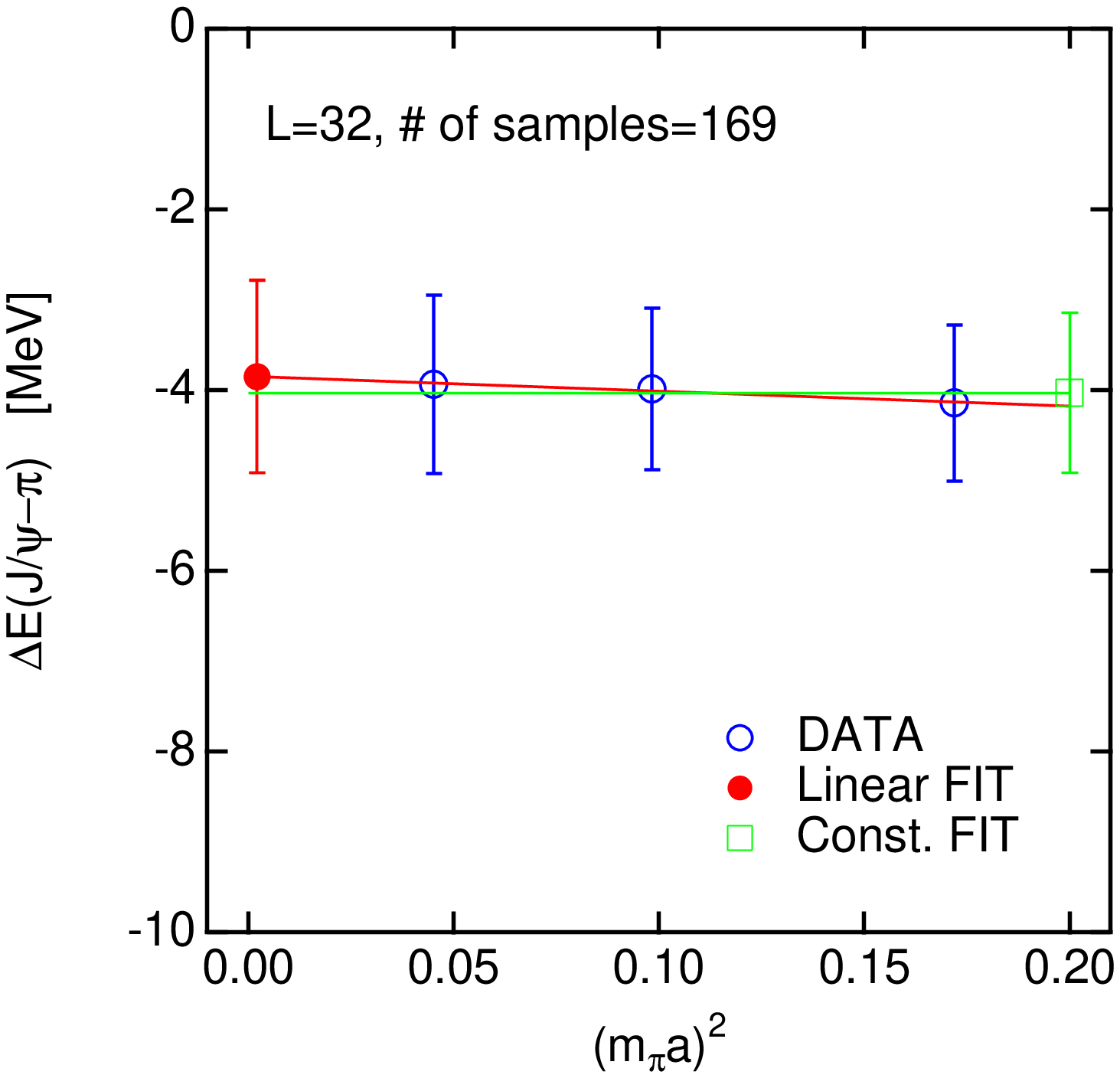, width=.48\textwidth}
\epsfig{file=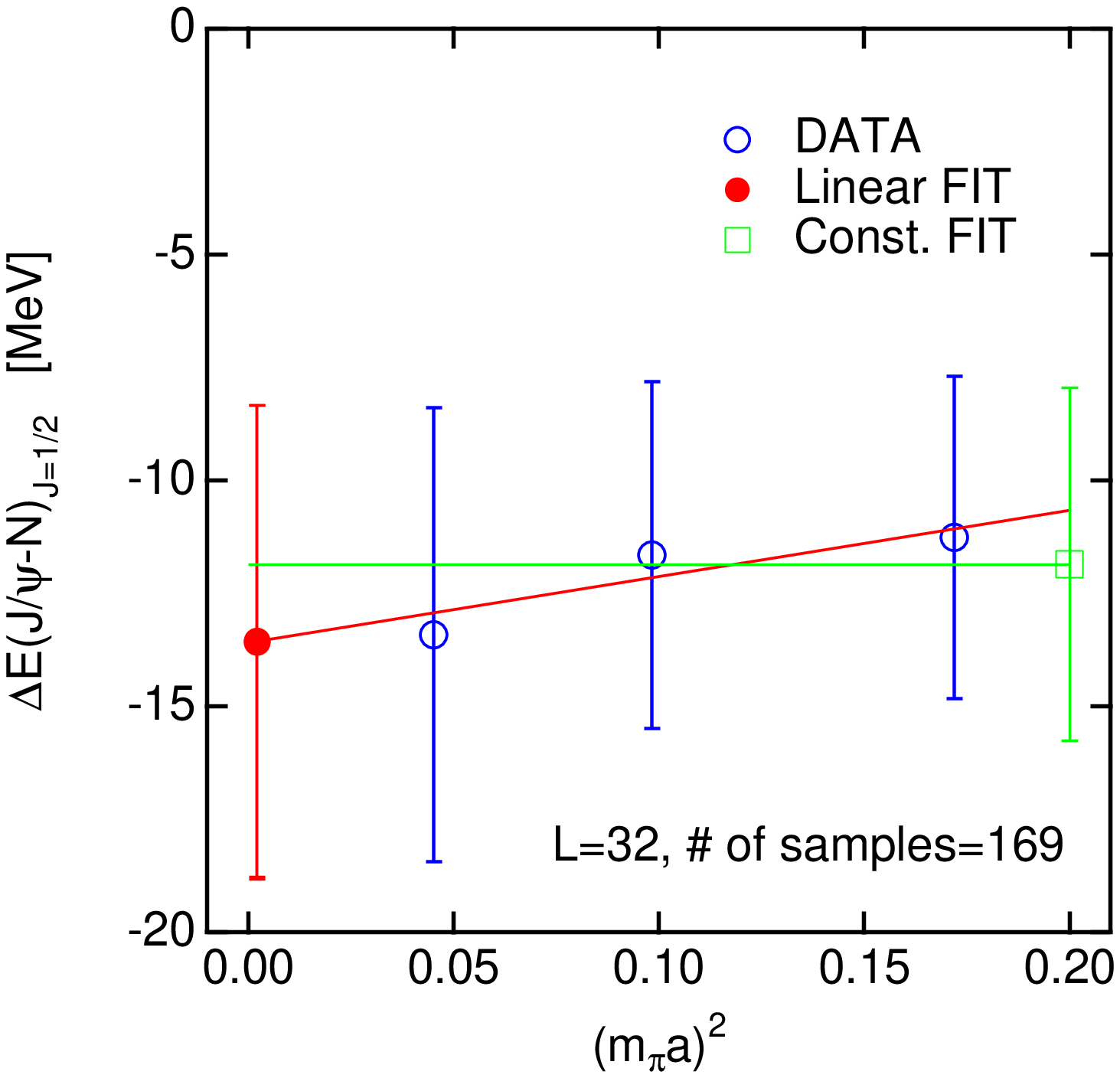, width=.48\textwidth}
\caption{The left (right) panel shows
the quark mass dependence of the energy shift for
the J/$\psi$-$\pi$ (J/$\psi$-$N$) interaction in $L=32$ lattice.
The horizontal axis is the pion mass squared $(m_\pi a)^2$ and 
the vertical axis is the energy shift $\Delta E$.
The open circles are the energy shift extracted from different $\kappa$.
The circles are the linear extrapolated points at the physical point.
The open squares are the points which are evaluated with an assumption
that energy shift is independent of quark mass.
}
\label{fig1}
\end{figure}
In Figure 1, we show the quark mass dependence of the energy shift $\Delta E$ in 
the J/$\psi$-$\pi$ channel (the left panel) and the J/$\psi$-$N$ channel
(the right panel).
The open circles are the energy shift extracted from the correlator ratio 
for different quark masses.
The circles and the open squares are the results of a
 linear fit in quark mass and of a simple average over the data with
 different quark masses, respectively.
The open squares are the estimate of $\Delta E$ without quark mass dependence as a reference.
In both channels, we found that the interactions are  attractive.

\begin{figure}
\epsfig{file=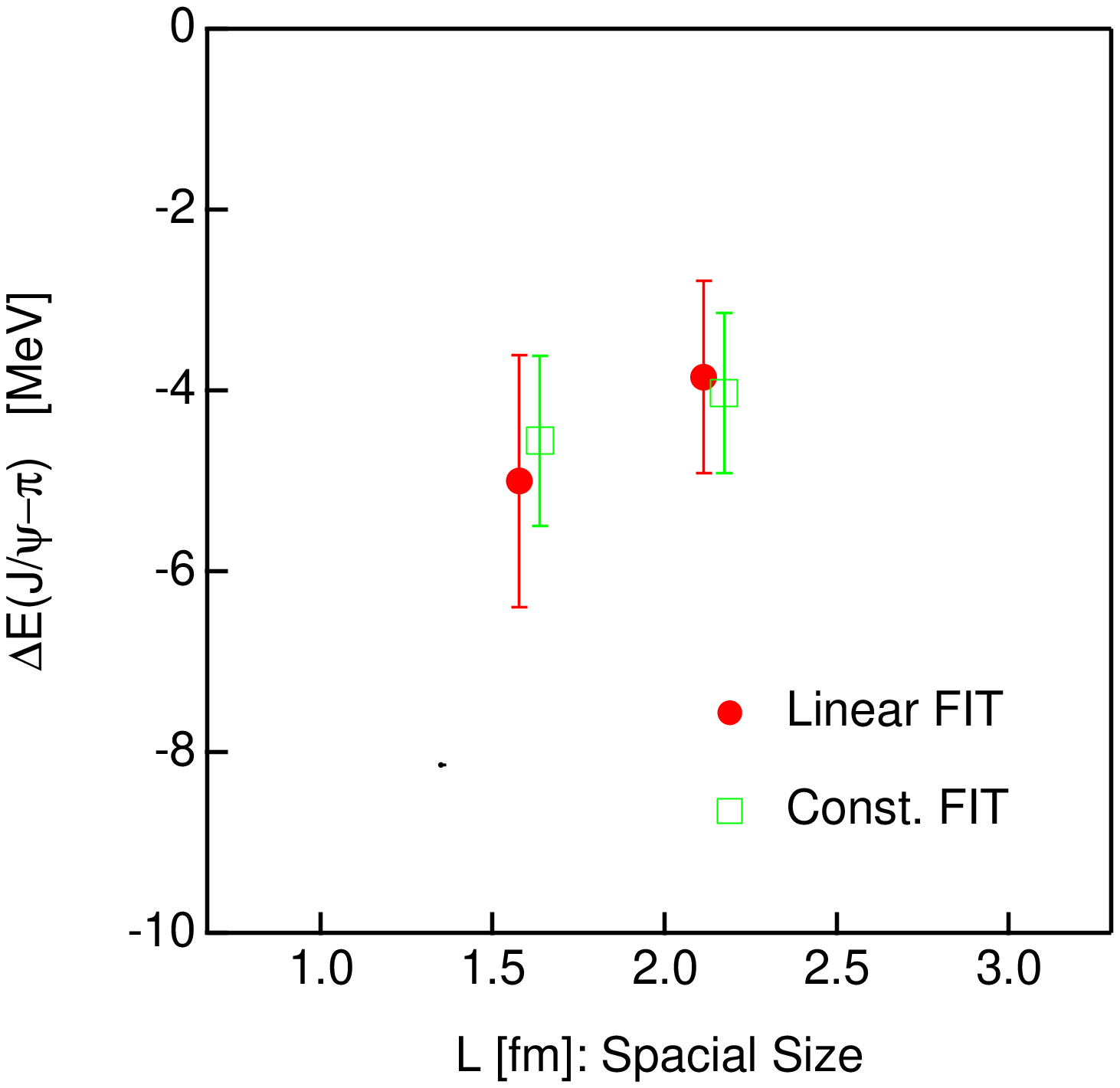, width=.48\textwidth}
\epsfig{file=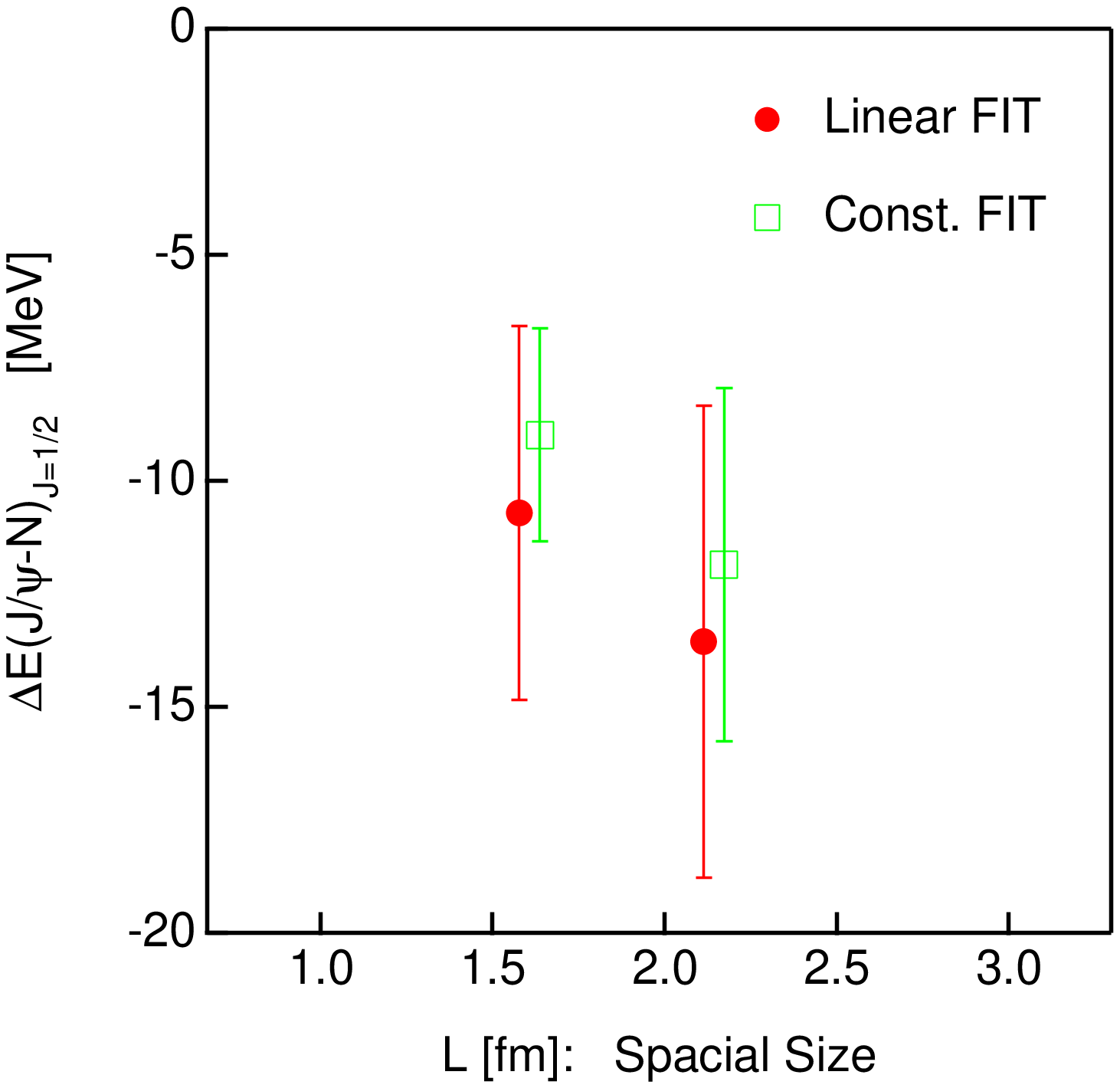, width=.48\textwidth}
\caption{The left (right) panel shows
the volume dependence of the energy shift from 
the J/$\psi$-$\pi$(J/$\psi$-$N$) interaction.
The horizontal axis is the spatial size $L$ and the vertical axis is the energy shift $\Delta E$.
The circles indicate the energy shifts which are assumed to 
have linear quark mass dependence extrapolated to the physical point.
The open squares show energy shifts estimated as if there is no quark mass dependence.
}
\label{fig2}
\end{figure}
 In Figure 2, we show the volume dependence of the  energy shift in 
the J/$\psi$-$\pi$ channel (the left panel) and the J/$\psi$-$N$ channel
 (the right panel).
 The circles are the energy shifts from linear quark mass extrapolation 
to the physical point and the open squares are the results of using the constant quark mass dependence.
Although the error bars are large in both channels, one can see the following tendency:
In the J/$\psi$-$\pi$ channel, the absolute value of $\Delta E$ decreases as $L$ increases.
On the other hand, $\Delta E$ for J/$\psi$-$N$ has opposite tendency.
However, to make firm conclusions on this, we need to increase statistics 
and also collect the data for larger $L$.
If it turns out to be true in high statistics data,
 one may conclude that J/$\psi$-$\pi$ channel is attractive without a bound state, 
while J/$\psi$-$N$ may have a bound state \cite{SS}.

\begin{figure}
\epsfig{file=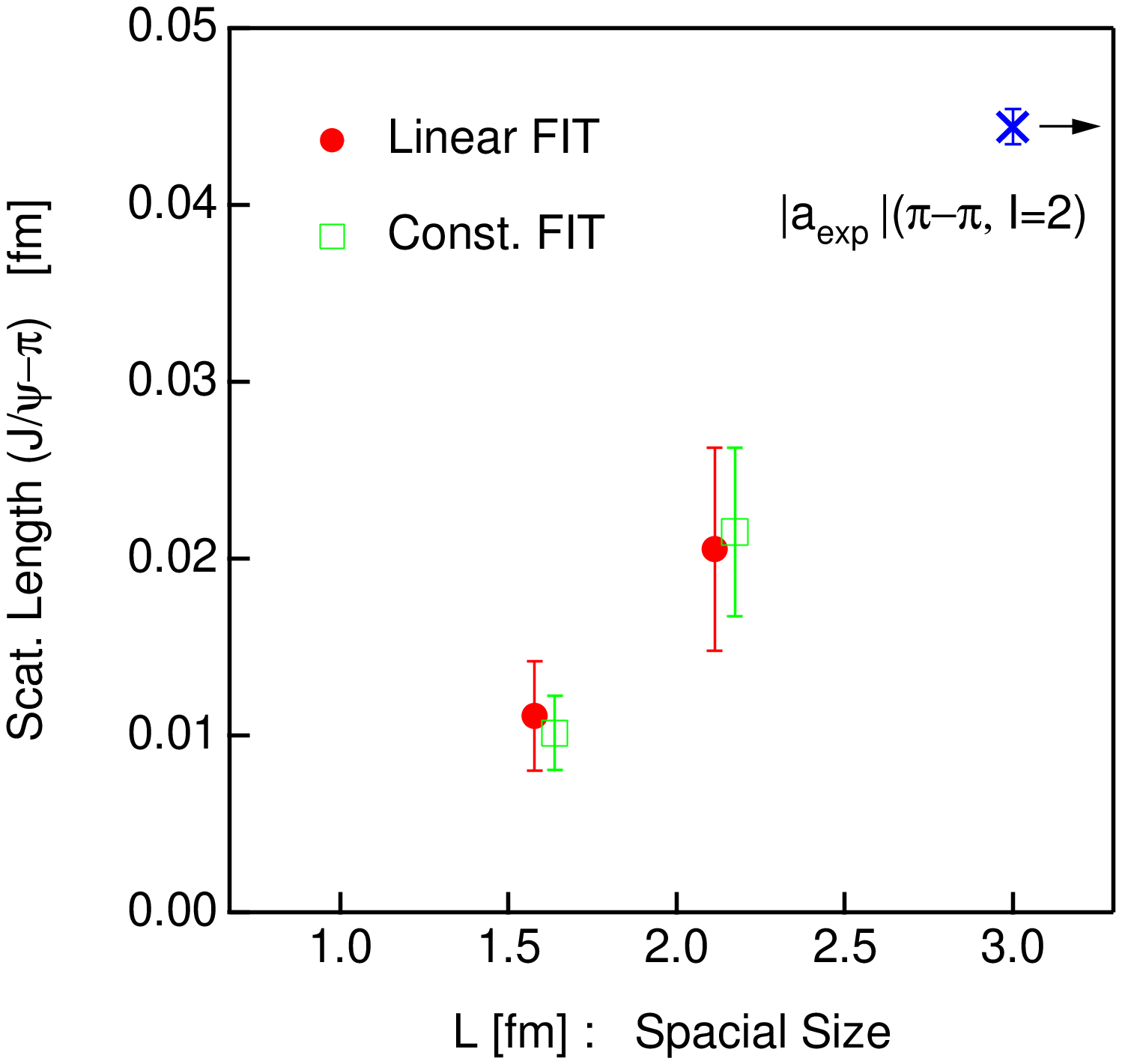, width=.48\textwidth}
\epsfig{file=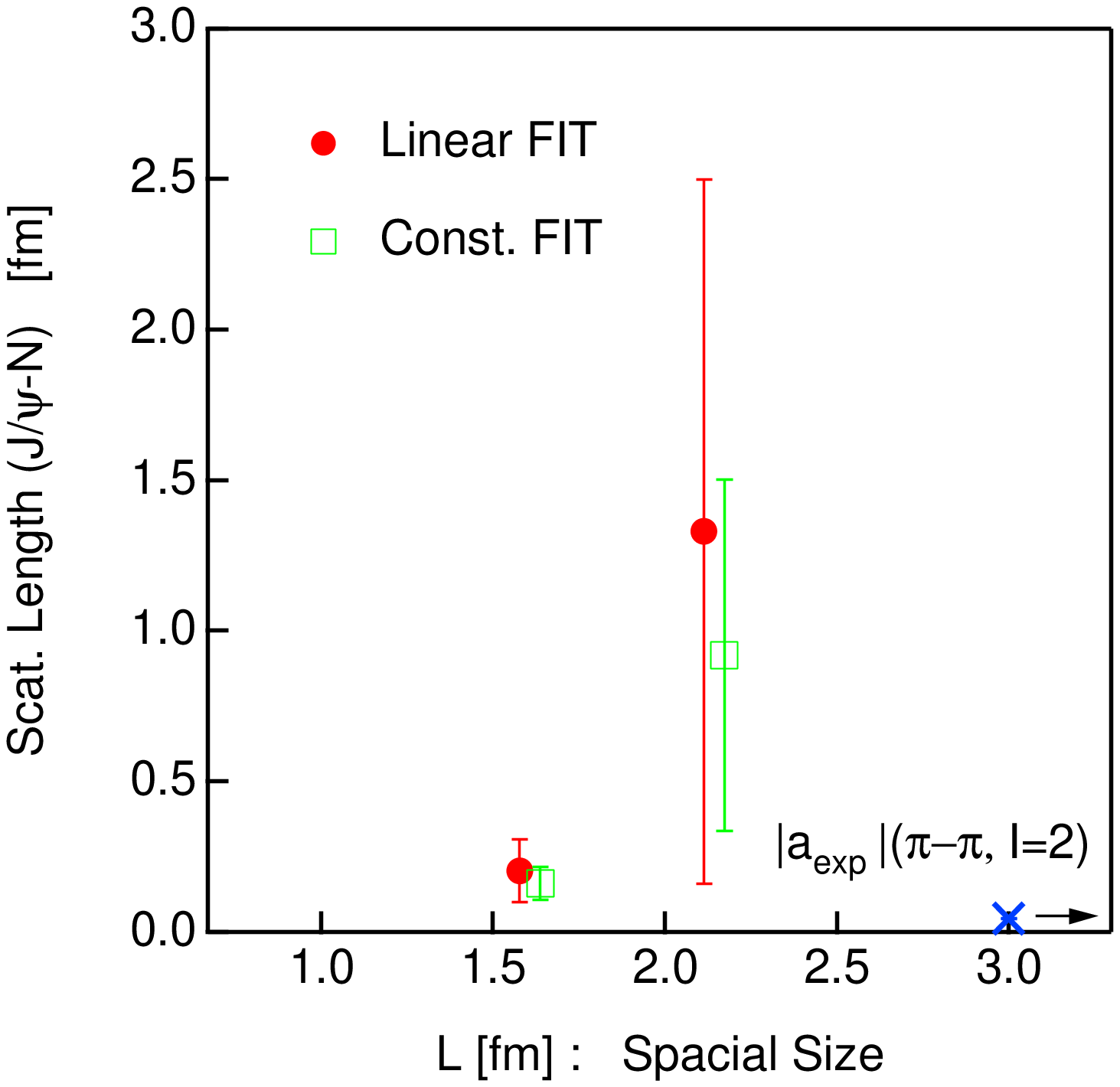, width=.47\textwidth}
\caption{
The left (right) panel shows
the volume dependence of the scattering length in 
the J/$\psi$-$\pi$ (J/$\psi$-$N$) channel.
The horizontal axis is the spatial size $L$ and the vertical axis is the S-wave scattering length $a_0$.
The circles (open squares) indicate the scattering lengths which are assumed 
to have linear (constant) quark mass dependence.
The crosses show the sign-flipped empirical value for the $\pi$-$\pi$ scattering lengths
in the $I=2$ channel as reference points.
}
\label{fig3}
\end{figure}
Finally, in Figure 3, we show the volume dependence of the scattering lengths in 
the J/$\psi$-$\pi$ (left panel) and the J/$\psi$-$N$ (right panel) channels.
The circles and the open squares are the results of linear quark mass extrapolation 
to the physical point and of fitting assumed constant quark mass dependence.
 To compare these absolute magnitudes of the J/$\psi$-hadron scattering length
 with that in the $I=2$ $\pi$-$\pi$ channel, 
we put the empirical value of the $\pi$-$\pi$ scattering length by crosses.
Note that $I=2$ $\pi$-$\pi$ scattering is repulsive and we show its
 absolute value  in the figure  for comparison.
The J/$\psi$-$\pi$ scattering length is negative and small compared to $\pi$-$\pi$.
This is partly because the size of J/$\psi$ is small than the pion, 
and partly because only the gluonic exchange is allowed in the J/$\psi$-$\pi$ case.
On the other hand, the scattering length for J/$\psi$-$N$ could be order of 
magnitude larger than J/$\psi$-$\pi$, although the error bar is still quite large.

\section{Summary}
In summary, we study the J/$\psi$-hadron scattering lengths by the  
quenched lattice QCD simulations.
We found attractive interactions in all J/$\psi$-$\pi$, J/$\psi$-$\rho$ and J/$\psi$-$N$
channels.  Furthermore, the J/$\psi$-$N$ scattering length is considerably larger than 
the J/$\psi$-meson scattering length.
Also, we found a sizable volume dependence of scattering lengths in all three channels.
There is an opposite tendency of the volume dependence 
 between  J/$\psi$-$\pi$ and J/$\psi$-$N$.
 There are several future problems to be examined further:
To study whether the attractive J/$\psi$-$N$ interaction could form a bound state,
we need to have better statistics and simulations with larger lattice volumes, which 
is now under way.
To confirm the validity of using the L\"uscher's formula,
we need to check whether the potential range is small enough in comparison to
 the lattice size.
Moreover, we need more careful analysis of inelastic contribution in our correlator ratio
 such as  the $D$-$\bar{D}$ contribution in the J/$\psi$-$\rho$ channel.

\subsection*{Acknowledgement}
This work was supported by the Supercomputer Projects No.110 (FY2004)
and No.125 (FY2005) of High Energy Accelerator Research
Organization (KEK). S.S. and T.H. were also supported by
  Grants-in-Aid of MEXT, No. 15540254 and No.15740137.

\end{document}